\documentclass[12pt]{article}

\usepackage{latexsym}
\usepackage{verbatim}
\usepackage{amssymb}

\catcode`\@=11
\def\marginnote#1{}
\newcount\hour
\newcount\minute
\newtoks\amorpm
\hour=\time\divide\hour by60
\minute=\time{\multiply\hour by60 \global\advance\minute by-\hour}
\edef\standardtime{{\ifnum\hour<12 \global\amorpm={am}%
        \else\global\amorpm={pm}\advance\hour by-12 \fi
        \ifnum\hour=0 \hour=12 \fi
        \number\hour:\ifnum\minute<10 0\fi\number\minute\the\amorpm}}
\edef\militarytime{\number\hour:\ifnum\minute<10 0\fi\number\minute}
\def\draftlabel#1{{\@bsphack\if@filesw {\let\thepage\relax
   \xdef\@gtempa{\write\@auxout{\string
      \newlabel{#1}{{\@currentlabel}{\thepage}}}}}\@gtempa
   \if@nobreak \ifvmode\nobreak\fi\fi\fi\@esphack}
        \gdef\@eqnlabel{#1}}
\def\@eqnlabel{}
\def\@vacuum{}
\def\draftmarginnote#1{\marginpar{\raggedright\scriptsize\tt#1}}
\def\draft{\oddsidemargin -.5truein
        \def\@oddfoot{\sl preliminary draft \hfil
        \rm\thepage\hfil\sl\today\quad\militarytime}
        \let\@evenfoot\@oddfoot \overfullrule 3pt
        \let\label=\draftlabel
        \let\marginnote=\draftmarginnote
   \def\@eqnnum{(\theequation)\rlap{\kern\marginparsep\tt\@eqnlabel}%
\global\let\@eqnlabel\@vacuum}  }

\def\preprint{\twocolumn\sloppy\flushbottom\parindent 1em
        \leftmargini 2em\leftmarginv .5em\leftmarginvi .5em
        \oddsidemargin -.5in    \evensidemargin -.5in
        \columnsep 15mm \footheight 0pt
        \textwidth 250mmin      \topmargin  -.4in
        \headheight 12pt \topskip .4in
        \textheight 175mm
        \footskip 0pt
        \def\@oddhead{\thepage\hfil\addtocounter{page}{1}\thepage}
        \let\@evenhead\@oddhead \def\@oddfoot{} \def\@evenfoot{} }

\def\titlepage{\@restonecolfalse\if@twocolumn\@restonecoltrue\onecolumn
     \else \newpage \fi \thispagestyle{empty}\c@page\z@ 
        \def\thefootnote{\fnsymbol{footnote}} }

\def\endtitlepage{\if@restonecol\twocolumn \else  \fi
        \def\thefootnote{\arabic{footnote}}
        \setcounter{footnote}{0}}  

\catcode`@=12
\relax

\def\bea{\begin{array}}
\def\bem{\begin{displaymath}}
\def\beq{\begin{equation}}

\def\eea{\end{array}}
\def\eem{\end{displaymath}}
\def\eeq{\end{equation}}

\def\Im{\mathop{\rm Im}}

\def\ov{\overline}

\def\Re{\mathop{\rm Re}}

\def\s2w{\sin^2 \theta_W}
\def\Tr{\mathop{\rm Tr}}

\relax
\def\crbig{\\\noalign{\vspace {3mm}}}
\def\bigint{{\displaystyle\int}}

\def\Dint{\bigint d^2\theta d^2\ov\theta\,}
\def\Fint{\bigint d^2\theta\,}
\def\Fbarint{\bigint d^2\ov\theta\,}
\relax

\DeclareMathAlphabet{\mathf}{OT1}{pzc}{m}{n}
\begin{document}
\topmargin-2.4cm
%
%
%
%
\begin{titlepage}
\begin{flushright}
CERN-PH-TH/2010-085 \\
May 2010
\end{flushright}
\vspace{1.0cm}

\begin{center}{\Large\bf
The Hypermultiplet with Heisenberg Isometry in \\ \vspace{3mm}
\boldmath{$N=2$} Global and Local Supersymmetry}
\vspace{1.0cm}

{\large\bf N. Ambrosetti$^1$, I. Antoniadis$^{2,3}$, J.-P. Derendinger$^1$
\\ 
\vspace{3mm}
and P. Tziveloglou$^{2,4}$}

\vspace{6mm}

{\small 
$^1$ Albert Einstein Center for Fundamental Physics \\
Institute for Theoretical Physics, Bern University \\
Sidlerstrasse 5, CH--3012 Bern, Switzerland \\

\vspace{2mm}

$^2$ Department of Physics, CERN - Theory Division \\
CH--1211 Geneva 23, Switzerland \\

\vspace{2mm}

$^3$ Centre de Physique Th\'eorique, UMR du CNRS 7644 \\
Ecole Polytechnique, F--91128 Palaiseau, France

\vspace{2mm}

$^4$ Department of Physics, Cornell University, Ithaca, NY 14853, USA 
}
\end{center}
\vspace{.6cm}

\begin{abstract}\noindent
The string coupling of $N=2$ supersymmetric compactifications of type II string theory on a 
Calabi-Yau manifold belongs to the so-called universal dilaton hypermultiplet, that has four real 
scalars living on a quaternion-K\"ahler manifold. Requiring Heisenberg symmetry, which is
a maximal subgroup of perturbative isometries, reduces the possible manifolds
to a one-parameter family that describes the tree-level effective action deformed by the only possible perturbative 
correction arising at one-loop level. A similar argument can be made at the level of 
global supersymmetry where the scalar manifold is hyper-K\"ahler. 
In this work, the connection between 
global and local supersymmetry is explicitly constructed, providing a non-trivial gravity decoupled 
limit of type II strings already in perturbation theory.
\end{abstract}


\end{titlepage}
\renewcommand{\theequation}{\arabic{section}.\arabic{equation}}

\setcounter{footnote}{0}
\setcounter{page}{0}
\setlength{\baselineskip}{.6cm}
\setlength{\parskip}{.2cm}
\newpage
%
%

\section{Introduction}\label{secintro}
\setcounter{equation}{0}

Type II superstring theories compactified on a Calabi-Yau threefold (CY$_3$) yield $N=2$ 
supersymmetry in four dimensions. Even if gauge symmetries are absent in perturbation theory, the 
study of the effective field theory plays an important role in moduli stabilization by fluxes, as well as in 
a more realistic framework, 
such as in the presence of orientifolds and D-branes. Besides supergravity 
fields, their massless spectra involve vector multiplets and hypermultiplets describing all K\"ahler class and 
complex structure deformations of the CY$_3$ manifold, as well as a universal hypermultiplet containing 
the string dilaton. Because of its special connection to the string coupling, the study of this universal 
hypermultiplet is an important problem per se. In the following, we restrict ourselves to this sector which 
also becomes exact in the particular case of a compactification with no K\"ahler class (complex structure) moduli in type IIB (IIA), or when these closed string moduli are fixed.

The dilaton hypermultiplet contains four real scalars parametrizing a quaternion-K\"ahler manifold, as 
required by $N=2$ supergravity~\cite{BW}. Two of them come from the NS-NS (Neveu-Schwarz) sector 
and correspond to the string dilaton (associated to the string coupling constant) and the universal axion, 
Poincar\'e dual of the antisymmetric tensor $B_{\mu\nu}$. The other two come from the R-R (Ramond) 
sector and are obtained from various $n$-form gauge potentials. On the IIB side, they 
correspond to another scalar (0-form) $C_0$ and the dual of the two-form $C_{\mu\nu}$. At the string 
tree-level, these four scalars live on the symmetric coset $SU(1,2)/SU(2)\times U(1)$ which is also 
K\"ahler~\cite{Fer}. Perturbative string corrections keep 
at least three isometries corresponding 
to the three independent shifts of the NS-NS axion and the R-R scalars, generating the Heisenberg 
algebra. Imposing just these isometries and the quaternion-K\"ahler structure, one finds that the only 
possible perturbative correction arises at one loop, destroying the K\"ahler structure of the 
manifold~\cite{oneloop}. This correction was computed in~\cite{oneloopold,oneloop} and was found to 
be proportional to the Euler number of the CY$_3$.

More precisely, in the context of IIB superstrings, the Heisenberg algebra is generated by a combination 
of the gauge symmetries of the two antisymmetric tensors $B_{\mu\nu}$ (NS-NS) and $C_{\mu\nu}$ 
(R-R) and of the shift symmetry of the R-R scalar $C_0$:
\beq
\label{Heis1}
\delta B_{\mu\nu} = 2 \, \partial_{[\mu}\Lambda_{\nu]} , \qquad\qquad
\delta C_{\mu\nu} = 2 \, \partial_{[\mu}\tilde \Lambda_{\nu]} + \lambda B_{\mu\nu} , \qquad\qquad
\delta C_0 = \lambda.
\eeq
As a consequence, the theory depends on the invariant three-forms
\beq
\label{Heis2}
H_{\mu\nu\rho} = 3\,\partial_{[\mu} B_{\nu\rho]}, \qquad\qquad
F_{\mu\nu\rho} = 3\,\partial_{[\mu} C_{\nu\rho]} - C_0 H_{\mu\nu\rho}
\eeq
and on $\partial_\mu C_0$. The Heisenberg algebra follows from
\beq
\label{Heis3}
[\delta_1 , \delta_2 ] \, C_{\mu\nu} = 2 \, \partial_{[\mu} \lambda_2\Lambda_{1\nu]}
- 2 \, \partial_{[\mu}  \lambda_1\Lambda_{2\nu]}.
\eeq
After reduction to four dimensions, the gauge symmetries imply that each tensor can be dualized 
into a scalar field with axionic shift symmetry. The third global symmetry (with parameter $\lambda$) 
combines then with the axionic shifts to realize again the Heisenberg algebra on three scalar fields. 

Indeed, one obtains three scalar fields $\varphi$, $\tau$ and $\eta = C_0$, with Heisenberg variations
\beq
\label{HeisScalar}
\delta\eta = c_X, \qquad \delta \varphi = c_Y, \qquad 
\delta \tau = c_Z - c_X\varphi\, .
\eeq
The scalars $\varphi$ and $\tau$ are Poincar\'e dual to $C_{\mu\nu}$ 
and $B_{\mu\nu}$, respectively. The duality relations are, schematically, 
$$
\partial_\mu\varphi \quad \sim \quad \epsilon_{\mu\nu\lambda\rho} F^{\nu\lambda\rho},
\qquad\qquad 
\partial_\mu\tau + \eta\,\partial_\mu\varphi \quad \sim \quad 
\epsilon_{\mu\nu\lambda\rho} H^{\nu\lambda\rho}\, .
$$
The algebra is $[X,Y] \sim Z$, with $Y$ and $Z$ generating the axionic shifts (with parameters $c_Y$ and 
$c_Z$), while $X$ generates the shift of the R-R scalar (with parameter $c_X$).
Notice that the central charge of the algebra is (depending on the representation) the gauge symmetry of the R-R tensor and the
axionic symmetry of $\tau$, dual to the NS-NS tensor.

Actually, as we will see later on, the Heisenberg algebra is extended by a fourth perturbative 
generator $M$ that rotates $X,Y$ and commutes also with the central charge$~Z$: 
\beq
\label{Mis}
\delta_M \eta=c_M\varphi\, ,\qquad\delta_M \varphi= -c_M \eta\, ,\qquad
\delta_M \tau = {c_M\over2}(\eta^2-\varphi^2).
\eeq
Equivalently, $M$ rotates the phase of the complex R-R scalar $\eta+i\varphi$.
As a result, the perturbative symmetry becomes the two-dimensional Euclidean group $E_2$ with central extension $Z$.

Imposing $N=2$ supersymmetry {\it and}\, Heisenberg symmetry is a powerful constraint.
In a previous work~\cite{AADT}, we have briefly analyzed its implications in global $N=2$ 
supersymmetry, where hypermultiplet scalars form a Ricci-flat hyper-K\"ahler manifold~\cite{AGF}. 
For a single hypermultiplet, we found a unique non-trivial hyper-K\"ahler space admitting a 
one-parameter deformation.  
In local supersymmetry where hypermultiplets live on quaternion-K\"ahler manifolds \cite{BW}, a
one-parameter family of solutions emerges \cite{oneloop} from the general analysis of Calderbank and 
Pedersen \cite{CP}. These similar results suggest a correspondence between the local and 
global cases which could be studied using a Ricci-flat limit of the 
quaternion-K\"ahler manifold preserving the Heisenberg algebra. This is the main goal of the present
article.

Taking the limit $\kappa\rightarrow0$ in a hypermultiplet theory coupled to $N=2$ supergravity is a
subtle problem. In contrast to the simplest case of $N=1$, $N=2$ supergravity imposes that 
the curvature of the hypermultiplet scalar manifold is proportional to
the gravitational coupling $\kappa^2$ and hence the curved, quaternion-K\"ahler Einstein metric
of the local hypermultiplet smoothly turns to a Ricci-flat hyper-K\"ahler metric. 
However, to obtain a non-trivial space, an appropriate limit must be defined, involving a new mass scale that should remain finite as Planck mass goes to infinity. This mechanism has
only been explicitly displayed for some particular cases, mostly using the quaternionic quotient 
method \cite{G1, G2}.

Thus, in this work, we establish the precise connection between the local and global $N=2$ 
supersymmetric actions of a single hypermultiplet with Heisenberg isometry, to explicitly obtain the 
gravity-decoupled limit. 
We first reconstruct the $N=2$ supergravity Lagrangian using the method of quaternionic quotient, 
starting with superconformal supergravity and imposing gauge conditions and constraints. We 
then define the zero-curvature limit that reduces the perturbative-corrected metric 
of the dilaton hypermultiplet to the non-trivial hyper-K\"ahler form found in~\cite{AADT}. It turns out 
that the deformation parameter corresponding to the one-loop correction plays a crucial role. 
Indeed, the zero-curvature limit of the tree level $SU(1,2)/SU(2)\times U(1)$ metric is trivial, leading 
to free kinetic terms. The presence of the one-loop parameter however allows for a non-trivial limit, 
giving rise to a hyper-K\"ahler metric that depends on a mass scale which remains finite as the 
four-dimensional Planck mass goes to infinity. At the same time, the value of the string coupling 
is tuned to a fixed value determined by the one-loop correction and can be made weak for large and positive Euler number of the CY$_3$ manifold, 
so that non-perturbative corrections remain suppressed while taking the gravity decoupled limit.

This paper is organized as follows. In Section~\ref{secglobal}, we recall the global construction of a four-dimensional hyper-K\"ahler manifold with the Heisenberg isometry in three formulations~\cite{AADT}: single-tensor which has the advantage of an off-shell $N=2$ supersymmetry formulation, scalar that provides a geometric description with a metric, and double tensor corresponding to the type IIB string basis\footnote{Note though that the basis of string vertex operators corresponds to the single-tensor representation~\cite{oneloop}.}. In Section~\ref{seclocal}, we first review the Calderbank-Pedersen metric with
Heisenberg symmetry and show that the latter is actually extended by a fourth generator which commutes with its central charge and rotates the other two. This generates a fourth perturbative isometry of the metric, as described above. We then rederive in  supergravity the quaternion-K\"ahler metric with the Heisenberg isometry~\cite{oneloop}, by taking an appropriate quaternionic quotient of the symmetric quaternion-K\"ahler space $Sp(2,4)/Sp(2)\times Sp(4)$ containing two hypermultiplets. Their reduction to one is achieved by gauging a symmetry corresponding to the central charge of the Heisenberg algebra. In Section~\ref{seclimit}, we take the zero-curvature limit, leading to the one parameter hyper-K\"ahler manifold with Heisenberg symmetry of Section~\ref{secglobal}. We thus find the correspondence of string fields in the rigid globally supersymmetric limit and we also discuss the coupling of the dilaton hypermultiplet to a D-brane where one of the two supersymmetries in non-linearly realized. Finally, Section~\ref{secfin} contains some concluding remarks.

\section{On the Heisenberg algebra and global supersymmetry}\label{secglobal}
\setcounter{equation}{0}

\subsection{Lagrangians}\label{secLagr}

Consider a $N=1$ globally supersymmetric theory with two superfields, a chiral $\Phi$ and a real linear $L$. 
It contains three real scalars, $\Re\phi=\Re\Phi|_{\theta=0}$, 
$\Im\phi=\Im\Phi|_{\theta=0}$, and $C=L |_{\theta=0}$, and
$L$ also depends on the curl of an antisymmetric tensor
$H_{\mu\nu\rho} = 3\,\partial_{[\mu} B_{\nu\rho]}$. The Lagrangian (up to two derivatives) is 
\beq
{\cal L} = \Dint {\cal H}(L, \Phi, \ov\Phi) + \Fint W(\Phi) + \Fbarint \ov W(\ov\Phi)\, .
\eeq
Besides the gauge invariance of $B_{\mu\nu}$ which does not act on the superfields, we also
impose a two-parameter global symmetry acting on $\Phi$ with variations
\beq
\label{Hsym1}
\delta\Phi = \alpha-i\beta .
\eeq
In this formulation, all three symmetries trivially commute. Nevertheless, in the version where 
$B_{\mu\nu}$ is dualized to a scalar, or in the version where $\Im\phi$ (for instance) is 
transformed into a second antisymmetric tensor, the three-parameter symmetry realizes a 
Heisenberg algebra acting either on three scalars according to Eq.~(\ref{HeisScalar}), as in the hypermultiplet formulation of IIB 
strings compactified to four dimensions, or on two tensors and one scalar according to Eqs.~(\ref{Heis1})
and (\ref{Heis3}). The Lagrangian compatible with the required symmetry (\ref{Hsym1}) has
\beq
\label{Hsym2}
{\cal H}(L,\Phi,\ov\Phi) = {\cal F}(L) + [AL+B]\Phi\ov\Phi,
\qquad
W(\Phi)=k\Phi,
\eeq
with an arbitrary function ${\cal F}(L)$ and real constants $A$ and $B$.\,\footnote{Of course, $B$ can be eliminated by a constant shift of $L$.} The constant $k$
generates a $C$--dependent potential $V= |k|^2/(AC+B)$ which does not admit a vacuum
if $A\ne0$. We take then $k=0$. 

The superfields $\Phi$ and $L$ provide an off-shell representation of the $N=2$ single-tensor
multiplet. On the $N=1$ Lagrangian, the condition for a second supersymmetry is 
\cite{LR}\footnote{The same conventions as in Ref.~\cite{AADT} are used. They slightly differ 
from Ref.~\cite{LR}.}
\beq
\label{Hsym3}
{\partial^2{\cal H}\over\partial L^2} + 2 {\partial^2{\cal H}\over\partial\Phi\partial\ov\Phi}=0,
\eeq
which in turn indicates that
\beq
\label{Hsym4}
{\cal F}_{N=2}(L) = -{A\over3} \, L^3 - BL^2.
\eeq
The same theory is given by
\beq
\label{Hsym4b}
\widehat{\cal F}_{N=2}(L) = -{1\over3A^2} \, (AL+B)^3.
\eeq
Hence, the $N=2$ theory compatible with complex shift symmetry of $\Phi$ is the sum
\beq
\label{Hsym5}
{\cal L}_{N=2} = \Dint \left[ A \left(-{1\over3}L^3 + L\Phi\ov\Phi\right) + B (-L^2+\Phi\ov\Phi)\right]
\eeq
of a trilinear interacting term and of a free term where the symmetry is trivial.
If canonical dimensions are assigned to $L$ and $\Phi$,  $A$ has dimension (mass)$^{-1}$ and $B$
is dimensionless.

Fur further use, we need the bosonic component expansion of this superfield theory.
Using
$$
\begin{array}{rcl}
L(x,\theta,\ov\theta)  &=& C + \theta\sigma^\mu\ov\theta\, v_\mu + {1\over4}\theta\theta\ov{\theta\theta} \Box C,
\qquad v_\mu = {1\over6}\epsilon_{\mu\nu\rho\sigma} H^{\nu\rho\sigma}, \qquad
H_{\nu\rho\sigma} = 3\,\partial_{[\nu}B_{\rho\sigma]}, 
\crbig
\Phi(x,\theta,\ov\theta) &=& \phi(x) - i\theta\sigma^\mu\ov\theta\, \partial_\mu \phi 
- \theta\theta f - {1\over4}\theta\theta\ov{\theta\theta} \Box \phi ,
\end{array}
$$
we obtain\footnote{The auxiliary field $f$ vanishes.}
\beq
\label{Hsymbos}
\begin{array}{rcl}
{\cal L}_{N=2, \, bos.} &=& (AC+B)\Bigl[ {1\over2}(\partial_\mu C)^2 
+ (\partial_\mu\phi)(\partial^\mu\ov\phi) + {1\over12} H^{\mu\nu\rho}H_{\mu\nu\rho} \Bigr]
\crbig
&& -{i\over12}A\, \epsilon^{\mu\nu\rho\sigma} (\ov\phi\,\partial_\mu\phi -\phi\,\partial_\mu\ov\phi)
H_{\nu\rho\sigma}.
\end{array}
\eeq
Since, $\partial_{[\mu}H_{\nu\rho\sigma]}=0$, the variation (\ref{Hsym1}) of $\phi$ induces 
a total derivative. Kinetic terms are positive if $AC+B > 0$. If $A\ne0$, $B$ can be eliminated by shifting 
$C$. The (shifted) field $C$ will be assumed strictly positive and the two options are an interacting, 
cubic theory with $A>0$ and $B=0$, or the free theory $A=0$, $B>0$.

We may then perform two supersymmetric duality transformations \cite{S} on theory (\ref{Hsym2}), either
turning the linear $L$ into a chiral $S$ or turning the chiral $\Phi$ into a second linear multiplet $L^\prime$.
The first transformation leads to
\beq
\label{Hsym7}
{\cal L} = \Dint \left[\widetilde{\cal F}({\cal Y}) + B\Phi\ov\Phi \right],
\eeq
where $\widetilde{\cal F}({\cal Y})$ is the Legendre transform of ${\cal F}(L)$ and the variable 
is\footnote{Notice that $\int d^2\theta d^2\ov\theta\,\Phi\ov\Phi 
= {1\over A}\int d^2\theta d^2\ov\theta\, {\cal Y} + {\rm derivative}$.} 
${\cal Y}=S+\ov S + A\Phi\ov\Phi$. Invariance of ${\cal Y}$ under shift symmetries (\ref{Hsym1}) 
requires a compensating variation of $S$:
\beq
\label{Hsym9}
\delta_H S = (\alpha\delta_X+\beta\delta_Y+\gamma\delta_Z)S 
= -A(\alpha+i\beta)\Phi + 2i \gamma, 
\eeq
where the axionic shift symmetry of $\Im S$ is dual to the gauge symmetry of $B_{\mu\nu}$, and the subscripts $X,Y,Z$ make clear the correspondence with the transformations (\ref{HeisScalar}). Indeed, since
\beq
\label{Hsym10}
[\delta_H^\prime,\delta_H] S \equiv -A(\alpha^\prime+i\beta^\prime) \delta_H\Phi 
+ A(\alpha+i\beta) \delta_H^\prime \Phi
= 2iA(\alpha^\prime\beta - \alpha\beta^\prime), 
\qquad
[\delta_H,\delta_H^\prime]\Phi = 0,
\eeq
the chiral theory has Heisenberg symmetry. Moreover, the theory (\ref{Hsym7}) has
another symmetry $M$ rotating the chiral superfield $\Phi$, as already mentioned in the Introduction (see Eq.~(\ref{Mis})).

For the $N=2$ single-tensor theory (\ref{Hsym5}), the dual hypermultiplet theory\footnote{With
positive K\"ahler metric.} is
\beq
\label{Hsym11}
{\cal L}_{N=2} = \Dint {\cal K}({\cal Y})=
{2\over3A^2}\Dint\left( A{\cal Y} + B^2 \right)^{3/2}.
\eeq
Eliminating some derivatives, the limiting case $A=0$ is a free theory. As required for a hyper-K\"ahler sigma-model, the determinant of the K\"ahler metric is constant (and positive).

A useful change of variable is
\beq
\label{Hsym12b}
\hat S = S - {A\over2}\Phi^2, \qquad\qquad 
{\cal Y}= \hat S + \ov{\hat S} + {A\over2}(\Phi+\ov\Phi)^2.
\eeq
and transformation (\ref{Hsym9}) becomes $\delta_H \hat S = -2A\alpha\Phi + 2i\gamma$.
With these variables, the transformations with parameters $\beta$ and $\gamma$ 
only act as shift symmetries of $\Im\Phi$ and $\Im\hat S$ respectively. In terms of variables ${\cal Y}$,
$\Im \hat S$, $\Re\Phi$ and $\Im\Phi$, one immediately deduces that the most general 
Heisenberg-invariant supersymmetric theory 
is of the form (\ref{Hsym7}).

Performing the second duality transformation of the chiral $\Phi$ into a linear $L^\prime$, 
always leads to the dual theory
\beq
\label{Hsym14}
{\cal L} = \Dint \left[{\cal F}(L) - {1\over2} { {L^\prime}^2\over AL+B} \right],
\eeq
with $\cal F$ given in Eq.~(\ref{Hsym4}). Expression (\ref{Hsym14})
is actually the most general $N=1$ Lagrangian for $L$ and $L^\prime$ with symmetry 
\beq
\label{Hsym13}
\delta L^\prime = \alpha(AL+B).
\eeq
This transformation, which links the two antisymmetric tensors in $L$ and $L^\prime$ as in variation
(\ref{Heis1}), forms with their respective gauge symmetries a Heisenberg algebra realized as in type 
IIB strings.

Instead of $\Im\Phi$, we could have chosen to dualize $e^{ia}\Phi$
for any phase $a$, since
$$
\Dint (AL+B)\Phi\ov\Phi 
= {1\over2}\Dint (AL+B) (e^{ia}\Phi + e^{-ia}\ov\Phi)^2
+{\rm derivative}.
$$
The result would be again theory (\ref{Hsym14}). This is a consequence of symmetry $M$, which is however fixed by the choice of dualization and
does not act on $L^\prime$.

\subsection{Hyper-K\"ahler metrics with Heisenberg symmetry}\label{secHeis}

The K\"ahler coordinates defined by $N=1$ chiral superfields $S$ and $\Phi$ are not necessarily 
the most appropriate to describe a hyper-K\"ahler manifold. There is
a `standard' set of coordinates used to describe hyper-K\"ahler 
metrics with shift isometries in the literature. For comparison purposes, 
we define
in this Subsection these coordinates in terms of our superfield components.

For any hyper-K\"ahler manifold with a shift symmetry, one can find 
coordinates in which the metric has the Gibbons-Hawking form \cite{GH}
\beq
\label{HK6}
ds^2 = f(\vec x) \, dx_i\, dx_i + f(\vec x)^{-1} (d\tau + \omega_i\, dx_i)^2 ,
\eeq
with condition $\vec\nabla\times\vec\omega = \vec\nabla f$.
Imposing the requirement of a Heisenberg symmetry acting according to 
\beq
\delta_H\, x_1 = \sqrt2\, \alpha, \qquad
\delta_H\, x_2 = -\sqrt2\, \beta, \qquad
\delta_H\, x_3 = 0, \qquad
\delta_H\, \tau = -\sqrt2\, \alpha \, x_2 + \gamma
\eeq
also defines $d\tau + x_1\, dx_2$ as the invariant derivative of $\tau$ and indicates that 
$\vec\omega = (0,x_1,0)$. The value of $f(\vec x)$ follows then from 
$\vec\nabla\times\vec\omega = \vec\nabla f$.
This last condition is invariant under $\vec\omega\rightarrow\vec\omega 
+ \vec\nabla\lambda(\vec x)$, for any gauge function $\lambda(\vec x)$. 
In turn, invariance of 
the metric requires the compensating transformation $\tau\rightarrow\tau-\lambda(\vec x)$.

From the $N=2$ K\"ahler potential (\ref{Hsym11}), the K\"ahler metric
can be written\footnote{From here on, we do not distinguish
chiral superfields $S$ and $\Phi$ and their lowest complex scalar components.}
\beq
\label{HK3}
\begin{array}{rcl}
ds^2 &=& {1\over2}(A{\cal Y}+B^2)^{-1/2}\left[ {1\over4} d{\cal Y}^2 
+ \Bigl(d\Im S +i\frac{A}{2}(\Phi\, d\ov\Phi - \ov\Phi \,d\Phi) \Bigr)^2\right]
\crbig
&& + (A{\cal Y}+B^2)^{1/2} \, d\Phi d\ov\Phi ,
\end{array}
\eeq
using coordinates $({\cal Y}, \Im S, \Re\Phi, \Im\Phi)$. 
The supersymmetric duality transformation
from $L$ to $S$ exchanges a real scalar $C=L|_{\theta=0}$, invariant under Heisenberg variations, and
$\Re S$ with variation (\ref{Hsym9}).
The Legendre transformation 
defines the change of variable from ${\cal Y}$ to $C$:
\beq
\label{HK4}
AC + B = \sqrt{A{\cal Y}+B^2}.
\eeq
Then, in terms of coordinates $(C, \Im S, \Re\Phi, \Im\Phi)$, the metric becomes
\beq
\label{HK5}
ds^2 = \displaystyle\frac{AC+ B}{2} \Bigl[ dC^2 + 2 \, d\Phi d\ov\Phi \Bigr]
\displaystyle + \frac{2}{(AC+ B)} \Bigl(d\tau + A\Re\Phi \,d\Im\Phi \Bigr)^2.
\eeq
This is the Gibbons-Hawking metric (\ref{HK6}) with 
$\vec x = (\sqrt2\Re\Phi, \sqrt2\Im\Phi, C)$ and
$$
\tau = {1\over2}\left(\Im S - A\Re\Phi\Im\Phi\right) = {1\over2}\Im \hat S .
$$
The function
\beq
f(\vec x) = {AC+B\over2}
\eeq
solves the hyper-K\"ahler condition 
$\vec\nabla\times\vec\omega = \vec\nabla f$ with 
$\vec\omega=(0,{A\over2}x_1,0)$.
Choosing for instance $\lambda= -{A\over2}x_1x_2$ turns then
$\vec\omega$ into $(-{A\over2}x_2, 0,0)$ and $d\tau + {A\over2}x_1dx_2$
into $d\tau - {A\over2}x_2dx_1$. Similarly, 
a rotation of $\Phi$
$$
\delta_M \, x_1= mx_2, \qquad\qquad \delta_M \, x_2 = -mx_1,
$$
which is compatible with the shift symmetry (\ref{Hsym1}), corresponds to 
$\lambda(\vec x)= {Am\over4}(x_2^2-x_1^2)$. 
It is the isometry $M$ of metric (\ref{HK5}).

The conclusion is that the Gibbons-Hawking Ansatz for the hyper-K\"ahler metric corresponds 
to coordinates where $\Re S$ is replaced by its Legendre dual $C$, which is also the lowest scalar
component of the linear superfield dual to $S$.

\section{The universal hypermultiplet in \boldmath{$N=2$} supergravity}\label{seclocal}
\setcounter{equation}{0}

Hypermultiplet scalars of $N=2$ supergravity live on $4n$--dimensional
quaternion-K\"ahler manifolds with holonomy included in $Sp(2n)\times Sp(2)$.
Supergravity requires that the curvature of these Einstein spaces is proportional to the gravitational 
coupling $\kappa^2$ \cite{BW}. Hence, the decoupling limit $\kappa\rightarrow0$ turns the hypermultiplet
manifold into a Ricci-flat hyper-K\"ahler space, as required by global $N=2$ supersymmetry \cite{AGF}. 
For a single hypermultiplet, or a four-dimensional quaternion-K\"ahler manifold, the defining condition 
on the holonomy is not pertinent since $Sp(2)\times Sp(2) \sim SO(4)$. The relevant condition is
then self-duality of the Weyl tensor.  

\subsection{The Calderbank-Pedersen metric with Heisenberg symmetry}\label{secCPH}

Calderbank and Pedersen \cite{CP} have classified all four-dimensional Einstein metrics with self-dual 
Weyl curvature and two commuting isometries. Using coordinates $(\rho, \eta, \varphi, \tau)$ with the
isometries acting as shifts of $\varphi$ and $\tau$, their metrics are written in terms of any single 
function $F(\rho,\eta)$ verifying
\beq
\label{CPFis1}
{\partial^2 F\over\partial\rho^2} + {\partial^2 F\over\partial\eta^2} 
= {3F\over 4\rho^2}.
\eeq
It is simple to see~\cite{oneloop} that metrics with Heisenberg symmetry are then obtained if $F$ does not depend 
on $\eta$, {\it i.e.} if\,\footnote{The metric does not make sense without the $\rho^{3/2}$ contribution 
to $F$ and the overall normalization of $F$ is a choice of coordinates. Our $\chi$ is
$\hat\chi$ in Ref.~\cite{oneloop}.} 
\beq
\label{CPFis2}
\sqrt\rho \, F(\rho) = {1\over2}[\rho^2 - \chi],
\eeq
with an arbitrary real parameter $\chi$. The Calderbank-Pedersen metric with Heisenberg symmetry 
(the CPH metric) reads then
\beq
\label{CP1}
ds^2_{CPH}=\frac{\rho^2+\chi}{(\rho^2-\chi)^2}(d\rho^2+d\eta^2+d\varphi^2)
+\frac{4\rho^2}{(\rho^2-\chi)^2(\rho^2+\chi)}(d\tau+\eta\, d\varphi)^2\,.
\eeq

The coordinate $\rho$ is positive, $\rho>0$, and positivity of the metric requires $\rho^2 + \chi>0$,
a stronger condition if $\chi$ is negative. 
It is an Einstein metric with negative curvature, and is K\"ahler only if $\chi=0$.
Notice that if $\chi\ne0$, the rescaling
$(\rho, \eta, \varphi, \tau)\rightarrow(|\chi|^{1/2}\rho, |\chi|^{1/2}\eta, |\chi|^{1/2}\varphi, |\chi|\tau)$
turns $\chi$ in metric (\ref{CP1}) into $\pm1$.
This is not true if we turn on string interactions, such as in the presence of D-branes where the dilaton, or equivalently the field $\rho$, couples to the Dirac-Born-Infeld (DBI) action in a non-trivial way (see Section~\ref{seclimit}). For this reason, we keep explicitly $\chi$ throughout the paper.
We may use a new coordinate $V=\rho^2$ with metric
\beq
\label{CP2}
ds^2_{CPH}=\frac{V+\chi}{(V-\chi)^2}\left({dV^2\over4V}+d\eta^2+d\varphi^2 \right)
+\frac{4V}{(V-\chi)^2(V+\chi)}\Bigl(d\tau+\eta\, d\varphi \Bigr)^2\,.
\eeq
The particular case $\chi=0$ has extended symmetry: it is the $SU(2,1)/SU(2)\times U(1)$ metric
with K\"ahler potential
\beq
\label{CP3}
K(\hat S,\ov{\hat S}, \Phi,\ov\Phi) = - \ln V, \qquad\qquad V = \hat S + \ov{\hat S} - (\Phi+\ov\Phi)^2,
\eeq
and with $\Phi = {1\over\sqrt2}(\eta+i\varphi)$, $\tau = -{1\over2}\Im \hat S$.

The CPH metric is invariant under four isometry variations acting on coordinates 
$(\eta,\varphi,\tau)$:
\beq
\label{CP4}
\begin{array}{rclrclrclrcl}
\delta_X\eta &=&  \sqrt2, \qquad
&\delta_Y\eta &=&  0, \qquad
&\delta_Z\eta &=&  0, \qquad
&\delta_M\eta &=& \varphi,
\crbig
\delta_X\varphi &=& 0, 
&\delta_Y\varphi &=& -\sqrt2,
&\delta_Z\varphi &=& 0, 
&\delta_M\varphi &=& -\eta,
\crbig
\delta_X\tau &=& -\sqrt2\,\varphi, 
&\delta_Y\tau &=& 0,
&\delta_Z\tau &=& 1, 
&\delta_M\tau &=& {1\over2}(\eta^2-\varphi^2 ).
\end{array}
\eeq
The non-zero commutators are
\beq
\label{CP5}
[X,Y] = 2Z, \qquad\qquad
[M,X] = Y, \qquad\qquad [M,Y]=-X.
\eeq
Hence, $X$, $Y$ and $Z$ generate the Heisenberg algebra and $Z$ is a central extension of a
two-dimensional euclidean algebra generated by $M$ (which rotates $\varphi$ and $\eta$), 
$X$ and $Y$ (which translate $\varphi$ and $\eta$). With these conventions,
\beq
\delta_H\,\Phi = (\alpha X + \beta Y + \gamma Z) \Phi = \alpha-i\beta,
\qquad\qquad
\delta_H\,\hat S =4\alpha\,\Phi - 2i\gamma
\eeq
and $V$ is invariant.

The metric (\ref{CP2}) appears in the one-loop-corrected Lagrangian of the universal hypermultiplet of type II strings, reduced to four dimensions,
with the NS-NS and R-R tensors dualized to scalars with shift symmetry
\cite{oneloop}. At one-loop order, the four-dimensional dilaton field is related to 
coordinate $V$ and parameter $\chi$ by
\beq
\label{dilaton}
e^{-2\phi_4} = V-\chi, \, \qquad\qquad \chi=-\chi_1, \qquad\qquad \chi_1 = {\chi_E\over12\pi},
\eeq
where $\chi_E$ is the Euler number of the internal CY$_3$ manifold.
The real number $\chi_1$ encodes the one-loop correction \cite{oneloop}. Notice that this
relation also indicates that $V-\chi=V+\chi_1>0$, which is stronger than $V=\rho^2 > 0$ if the Euler 
number is negative $(\chi>0$). Since positivity of the CPH metric also requires $V+\chi>0$ if $\chi<0$, the
domain of $V$ is naturally restricted to $V>|\chi|$. 

The R-R scalar is
\beq
\label{C0}
C_0 \equiv \eta\, ,
\eeq
and is shifted by symmetry $X$. Finally,
Poincar\'e duality gives the following equivalences
$$
\begin{array}{rcl}
d\varphi \quad  &\sim& \quad F_3 = dC_2 - \eta \, dB_2 ,
\crbig
d\tau + \eta\,d\varphi  \quad &\sim& \quad 
H_3 = d B_2 .
\end{array}
$$
In the scalar version, the central charge is the shift $Z$ of $\tau$ (related to the NS-NS tensor $B_2$)
while in the two-tensor version, it is the gauge variation of the (R-R) tensor $C_2$.
Writing $\eta$ and $\varphi$ in a complex $\Phi$ is conventional: we 
always use
$$
\Phi = {1\over\sqrt2}(\eta + i \varphi).
$$

In the previous Section, we found a unique four-dimensional hyper-K\"ahler manifold with Heisenberg 
symmetry. It also admits the fourth isometry $M$ rotating $\Phi$.
In the quaternion-K\"ahler case, the theorem of Calderbank-Pedersen \cite{CP} leads 
then to a very similar uniqueness conclusion. We will see how these two results are connected 
when taking an appropriate zero-curvature limit. But we first want to obtain the $N=2$ supergravity 
coupling of the universal hypermultiplet on the CPH manifold.

\subsection{Coupling to \boldmath{$N=2$} supergravity}\label{secsugra}

There are different methods to construct hypermultiplet couplings to $N=2$ supergravity. The
simplest procedure, which is however not the most general, is to use hypermultiplets coupled to 
local $N=2$ superconformal symmetry \cite{N=2conf} and to perform a quaternionic quotient 
\cite {G1, G2} using supplementary hypermultiplet(s) and non-propagating vector multiplet(s). 
In this Section, we use this procedure to obtain the supergravity theory of the one-loop-corrected 
dilaton hypermultiplet.

Related constructions, using more general but also more complicated methods, can be found in 
Ref.~\cite{ARV}, in the language of projective superspace or in 
Ref.~\cite{CIV}, using harmonic superspace.

Conformal $N=2$ supergravity is the gauge theory of $SU(2,2|2)$, which has a 
$SU(2)_R\times U(1)_R$ $R$--symmetry with non-propagating gauge fields. Pure 
Poincar\' e $N=2$ supergravity is obtained from the superconformal coupling of one propagating 
vector multiplet\footnote{Its gauge field is the graviphoton.} (which may be charged under $U(1)_R$) 
and one hypermultiplet (charged under $SU(2)_R$) by gauge-fixing of the extraneous symmetries.
These two multiplets include in particular the compensating fields used in the gauge-fixing 
to the Poincar\'e theory.

For the superconformal construction of our particular hypermultiplet sigma-model, we also need a
physical hypermultiplet, with positive kinetic metric, to describe the dilaton multiplet. In addition,
for the quaternionic quotient, we need a non-propagating vector multiplet with gauge
field $W_\mu$, gauging a specific generator $T$ to be discussed below, and,
since the elimination of the algebraic vector multiplet involves three constraints and one gauge choice 
on scalar fields, we also need a third non-physical hypermultiplet. Its kinetic metric can have a positive
or negative sign, depending on the constraints induced  by the choice of $T$.
Hence, we need to consider the $N=2$ superconformal theory of two vector multiplets and three 
hypermultiplets. The superconformal hypermultiplet scalar sector has then an
`automatic' $Sp(2,4)$ global symmetry in which the gauge generator $T$ 
of the quaternionic quotient is chosen.

\subsection{\boldmath{$Sp(2,4)$}}\label{secsl24}

In the following, we consider three hypermultiplets coupled to (superconformal) $N=2$ supergravity.
One (compensating) hypermultiplet has negative signature, the physical hypermultiplet has positive 
signature, the third hypermultiplet, associated to the non-propagating vector multiplet, may have a positive 
or negative signature, depending on the constraints applied to the scalar fields. In any case, we are considering $Sp(2,4)$--invariant supergravity couplings 
of $N=2$ hypermultiplets. 

The hypermultiplet scalars are $A_i^\alpha$, with $SU(2)_R$ index $i=1,2$ and $Sp(2,4)$ index 
$\alpha=1,\ldots,6$. They transform in representation $({\bf6},{\bf2})$ of $Sp(2,4)\times SU(2)_R$. 
Their conjugates are\footnote{We follow the conventions of the second paper of Ref.~\cite{N=2conf}.}
\beq
\label{conf2}
A^i_\alpha = (A_i^\alpha)^* = \epsilon^{ij}\rho_{\alpha\beta} A^\beta_j
\eeq
with $\rho^{\alpha\beta}\rho_{\beta\gamma} = - \delta^\alpha_\gamma$ and 
$\epsilon^{ij}\epsilon_{jk} = - \delta^i_k$. 
We choose the $Sp(2,4)$--invariant metric as 
\beq
\label{conf3}
\rho = I_3 \otimes i\sigma_2 = \left( \begin{array}{cc} 0 & I_3  \\ -I_3  & 0 \end{array}\right)
\eeq
and we use
\beq
\label{conf4}
d = \left(\begin{array}{cc} \eta&0\\0&\eta\end{array}\right),
\qquad\qquad \eta = {\rm diag}( -1 , 1 , -1 ), \qquad\qquad
\rho \,d\, \rho = -d.
\eeq
In our choice of $\eta$, direction 1 corresponds to the superconformal compensator, direction
2 to the physical hypermultiplet and our choice of quaternionic quotient will require a negative metric in direction 3; otherwise, our construction does not work. 
On scalar fields, $Sp(2,4)$ acts according to
\beq
\label{conf5}
\delta A_i^\alpha = g \,{t^\alpha}_\beta A^\beta_i, \qquad\qquad
\delta A^i_\alpha = g \,{t_\alpha}^\beta A_\beta^i, \qquad\qquad
{t_\alpha}^\beta = -\rho_{\alpha\gamma} \,{t^\gamma}_\delta\, \rho^{\delta\beta}.
\eeq
Since relation (\ref{conf2}) also implies ${t_\alpha}^\beta = ({t^\alpha}_\beta)^*$, the choice
(\ref{conf3}) and the invariance of $d^\alpha_\beta A_\alpha^i A^\beta_i$ lead to
\beq
\label{conf7}
t = \left( \begin{array}{cc}  U & \eta Q \\  -\eta Q^* & U^* \end{array}  \right),
\qquad\qquad U^\dagger = -\eta U \eta, \qquad
Q = Q^\tau, \qquad t^\dagger = - d \, t \, d.
\eeq
This is an element of $Sp(2,4)$: $U$ generates the $U(1,2)$ subgroup (9 generators) and
$Q$ (12 generators) generates $Sp(2,4)/ U(1,2)$.
The $(2\times2)$ matrix $A^\dagger \, d \, t \, A$, with matrix elements $A^i_\alpha d^\alpha_\beta {t^\beta}_\gamma A^\gamma_j$, is antihermitian, as required by gauge invariance of $A^\dagger d A$,
and traceless. 

\subsection{The Heisenberg subalgebra of \boldmath{$SU(1,2)$} and $Sp(2,4)$}\label{secHeislocal}

At string tree-level, the universal hypermultiplet of the dilaton in type II strings lives,
when formulated in terms of four real scalars, on the quaternion-K\"ahler and K\"ahler 
manifold $SU(1,2) / SU(2)\times U(1) = U(1,2)/U(2)\times U(1)$. Since $U(1,2)=SU(1,2)\times
U(1)_0$ is maximal in $Sp(2,4)$, $Sp(2,4)$ has a unique generator commuting with $SU(1,2)$:
the generator of $U(1)_0$. At one-loop however, the isometry is reduced and includes the Heisenberg 
algebra which is known to be a subalgebra of $SU(1,2)$. We need to find the most general 
generator $T$ of $Sp(2,4)$ which commutes with a Heisenberg subalgebra. In the following 
subsections, we will perform the quaternionic quotient construction induced by the gauging 
of $T$. 

Since elements $U$ of the $U(1,2)$ algebra verify $U^\dagger = -\eta \, U \, \eta$ and we
have chosen $\eta = {\rm diag}( -1 , 1 , -1 )$, a generic $U$ is
\beq
U = \left( \begin{array}{ccc} ia & A & B \\ \ov A & ib & C \\ -\ov B & \ov C & ic \end{array} \right),
\eeq
with $a$, $b$, $c$ real, $A$, $B$, $C$ complex and elements of $SU(1,2)$ are traceless.
On a three-dimensional complex vector, $U(1,2)$ variations are $\delta A = UA$.

We may define the Heisenberg subalgebra as the $U(1,2)$ transformations leaving 
$A_1-A_2$ invariant:
$(\delta_H A)_1 - (\delta_H A)_1= (UA)_1 - (UA)_2 =0$. The transformations acting on $A_1$ and $A_2$ are generated by the following three elements 
\beq
X = \left( \begin{array}{ccc} 0 & 0 & 1 \\ 0 & 0 & 1 \\ -1 & 1 & 0 \end{array} \right), \qquad
Y = \left( \begin{array}{ccc} 0 & 0 & i \\ 0 & 0 & i \\ i & -i & 0 \end{array} \right), \qquad
Z = \left( \begin{array}{ccc} i & -i & 0 \\ i & -i & 0 \\0 & 0 & 0 \end{array} \right)
\eeq
which verify 
\beq
0 = XZ = ZX = YZ = ZY=Z^2, \quad XY = - YX = Z, \quad
X^2 = Y^2 = iZ.
\eeq
The Heisenberg algebra 
\beq
\label{Heisdef}
[X,Y] = 2 Z , \qquad\qquad [X,Z] = [Y,Z] = 0
\eeq
is then realized as a subalgebra of $SU(1,2)$, with variations
\beq
\label{Heisvar}
\delta_H\, A = (\alpha X + \beta Y + \gamma Z) \, A
= \left(\begin{array}{ccc}  i\gamma & -i\gamma & \alpha + i\beta \\
i\gamma & -i\gamma & \alpha + i\beta \\ -\alpha + i\beta & \alpha - i\beta & 0
\end{array}\right)
\left(\begin{array}{c} A_1 \\ A_2 \\ A_3 \end{array}\right)
\eeq
in the fundamental representation.
Since $Z$ is a central charge of the Heisenberg algebra, we are interested in the
elements of $U(1,2)$ which commute with $Z$. They form an algebra generated by
five elements, $U_0$, $M$, $X$, $Y$ and $Z$, with
\beq
U_0 = i I_3, \qquad\qquad
M = i\left( \begin{array}{ccc} 1 & 0 & 0 \\ 0 & 1 & 0 \\ 0 & 0 & -2 \end{array} \right)
\eeq
($U_0$ generates the abelian factor of $U(1,2)=SU(1,2)\times U(1)_0$).
Besides the Heisenberg algebra generated by $X,Y,Z$, we also have
\beq
[ M, X ] = 3 Y, \qquad [ M, Y ] = -3 X
\eeq
and $M$ generates a rotation of $(X,Y)$ leaving $X^2 + Y^2 = 2iZ$ invariant:
$[M,X^2+Y^2] = 2i[M,Z]=0$.

One then easily checks that the most general $U(1,2)$ generator which commutes with the 
Heisenberg algebra generated by $X,Y,Z$ is proportional to
\beq
\label{Tis1}
\widehat T = U_0 + \chi\, Z =
i \left( \begin{array}{ccc} 1+\chi & -\chi & 0 \\ \chi & 1-\chi & 0 \\0 & 0 & 1 \end{array} \right), 
\qquad\qquad U_0 = iI_3,
\eeq
where $\chi$ is an arbitrary real number. If $\chi=0$, $\widehat T=U_0$ commutes with the whole 
$U(1,2)$. If $\chi\ne0$, $\widehat T$ commutes with the Heisenberg algebra supplemented 
by $U_0$ and $M$.
The extension to $Sp(2,4)$ is straightforward. Requiring that 
\beq
\label{Tis2}
T =  \left( \begin{array}{cc} \hat T & 0  \\  0 & \hat T^* \end{array}  \right)
\eeq
in $Sp(2,4)$ commutes with an element of $Sp(2,4)/U(1,2)$ corresponds to find a (non\-zero)
symmetric matrix $Q$ in Eq.~(\ref{conf7}) such that $\hat T^\dagger Q$ is also antisymmetric, 
which is impossible.\footnote{This would not be true for $\hat T=Z$, which commutes with a larger 
subalgebra of $Sp(2,4)$. The $U_0$ component is necessary.} Hence,
$T$ is also the most general generator in $Sp(2,4)$ which commutes with the Heisenberg algebra generated by $X$, $Y$ and $Z$ in $SU(1,2)$. It actually commutes with $X$, $Y$, $Z$, $M$ 
and $U_0$.

\subsection{\boldmath{$N=2$} supergravity scalar Lagrangian}\label{secsugra2}

To construct the scalar kinetic metric, 
the relevant terms of the $N=2$ conformal supergravity Lagrangian are \cite{N=2conf, G1, G2}
\beq
\label{conf9}
\begin{array}{rcl}
e^{-1}{\cal L} &=& d^\alpha_\beta (D_\mu A^\beta_i) (D^\mu A_\alpha^i)
+ (g \, d^\alpha_\beta \, A^i_\alpha {T^\beta}_\gamma A^\gamma_k \, Y^k_i + {\rm c.c.})
\crbig
&& +{1\over6} R ( - X_0\ov X_0 + d^\alpha_\beta A_\alpha^i A^\beta_i )
+ d (X_0 \ov X_0  + {1\over2} d^\alpha_\beta A_\alpha^i A^\beta_i).
\end{array}
\eeq
The complex scalar $X_0$ is the partner of the graviphoton, $Y^i_j$,  $Y^i_i=0$, is the triplet of real
auxiliary scalars in the non-propagating vector multiplet with gauge field $W_\mu$ used in the
quaternionic quotient. The covariant derivatives are
\beq
\label{conf10}
\begin{array}{rcl}
D_\mu A_i^\alpha &=& \partial_\mu A_i^\alpha - g^\prime W_\mu {T^\alpha}_\beta A^\beta_i
- g {V_{\mu i}}^j A_j^\alpha,
\crbig
D_\mu A^i_\alpha &=& \partial_\mu A^i_\alpha - g^\prime W_\mu {T_\alpha}^\beta A_\beta^i
- g {{V_\mu}^i}_j A^j_\alpha,
\end{array}
\eeq
where $g$ and $g^\prime$ are $SU(2)_R$ and $U(1)_T$ coupling constant.
The (anti-hermitian) $SU(2)$ gauge fields ${V_{\mu\, i}}^j$, ${V_{\mu\, i}}^i=0$, and the 
real auxiliary scalar $d$ belong to the multiplet of superconformal gauge fields:
$$
{V_{\mu\, i}}^j = {i\over2}V_\mu^x {(\sigma^x)_i}^j, \qquad\qquad
{{V_\mu}^i}_j = \epsilon^{ik}\epsilon_{jl} {V_{\mu\, k}}^l  = ({V_{\mu\, i}}^j )^*.
$$
We will commonly use a matrix notation, with a $6\times2$ complex matrix $A$ and its $2\times6$ 
conjugate $A^\dagger$ replacing $A^\alpha_i$ and $A_\alpha^i$. 
Condition (\ref{conf2}) implies that $A$ contains six complex components only. It also implies,
in particular, that $A^\dagger dA = {1\over2}\Tr(A^\dagger dA) \, I_2$. Since 
$V_\mu=-V^\dagger_\mu$, the Lagrangian and the derivatives read
\beq
\label{conf10b}
\begin{array}{rcl}
e^{-1}{\cal L} &=& \Tr (D_\mu A^\dagger)d(D^\mu A) +  g\Tr Y  A^\dagger d\,TA + {\rm c.c.}
\crbig
&& +{1\over6} R ( - X_0\ov X_0 + \Tr A^\dagger dA )
+ d( X_0\ov X_0 + {1\over2}\Tr A^\dagger dA );
\crbig
D_\mu A &=& \partial_\mu A - g^\prime W_\mu TA - g A V_\mu,
\crbig
D_\mu A^\dagger&=& \partial_\mu A^\dagger - g^\prime W_\mu A^\dagger T^\dagger
+ g V_\mu A^\dagger.
\end{array}
\eeq
Constraints are obtained from the elimination of the auxiliary fields and from the gauge-fixing  
of dilatation symmetry in the Poincar\'e theory:
\begin{itemize}
\item 
Einstein frame gauge-fixing condition and $d$ auxiliary field equation:
\beq
\label{conf11}
X_0\ov X_0 = {1\over\kappa^2}, \qquad\qquad
\Tr A^\dagger dA= - {2\over\kappa^2}.
\eeq
The second condition is invariant under $SU(2)_R$ and $Sp(4,2)$. With an $SU(2)$ gauge choice, 
it allows to eliminate four scalar 
fields and would lead to the $Sp(4,2)/Sp(4)\times Sp(2)$ sigma-model.
\item
Auxiliary fields $Y^i_j$:
\beq
\label{conf12}
A^\dagger d\,TA = 0. 
\eeq
Since this $2\times2$ matrix is traceless and antihermitian, 
these conditions eliminate three scalars and the associated abelian gauge invariance 
removes a fourth field. 
\end{itemize}
The $SU(2)_R$ gauge fields ${V_{\mu i}}^j$ and the abelian $W_\mu$ have then algebraic 
field equations:
\begin{itemize}
\item
Gauge field $W_\mu$, associated with generator $T$:
\beq
\label{conf13}
W_\mu = {\Tr(\partial_\mu A^\dagger d\,T A - A^\dagger d \, T \partial_\mu A) 
\over 2 g^\prime \Tr( A^\dagger T^\dagger d\,TA)}.
\eeq
\item
$SU(2)_R$ gauge fields ${V_{\mu \, i}}^j$:
\beq
\label{conf14}
V_\mu = -{ \partial_\mu A^\dagger d\,A - A^\dagger d\,\partial_\mu A \over 
g\Tr (A^\dagger d A)} .
\eeq
According to the second Eq.~(\ref{conf11}), the denominator is $- 2g/ \kappa^2$.
\end{itemize}
At this point, the scalar kinetic Lagrangian in theory (\ref{conf9}) reduces to
\beq
\label{conf15}
\begin{array}{rcl}
e^{-1}{\cal L} &=& 
e^{-1}({\cal L}_{kin.} + {\cal L}_{T} + {\cal L}_{SU(2)} )
\crbig
&=&
\Tr (\partial_\mu A^\dagger) d (\partial^\mu A)
- {g^\prime}^2 \Tr(A^\dagger T^\dagger d\,TA) W^\mu W_\mu 
- {g^2\over\kappa^2} \Tr (V^\mu V_\mu).
\end{array}
\eeq
The scalar fields are submitted to constraints (\ref{conf11}) and (\ref{conf12}) and the gauge fields
$W_\mu$ and ${V_{\mu\, i}}^j$ are defined by their field equations (\ref{conf13}) and (\ref{conf14}). 

To study the constraints (\ref{conf11}) and (\ref{conf12}) for our specific choice (\ref{Tis1})
and (\ref{Tis2}) of gauged generator $T$, we introduce two three-component complex vectors:
\beq
\label{conf16}
A^\alpha_i = \left( \begin{array}{cc}  \vec A_+ & \vec A_- \\
-\vec A_-^* & \vec A_+^*  \end{array} \right),  \qquad\qquad
A_\alpha^i = \left( \begin{array}{cc}  \vec A_+^* & \vec A_-^* \\
-\vec A_- & \vec A_+  \end{array} \right),
\eeq
verifying the reality condition (\ref{conf2}). On each doublet $A_{+a}$, $A_{-a}$, $a=1,2,3$, act two 
different $SU(2)$ groups. Firstly, the superconformal $SU(2)_R$ acts on $\pm$ indices. Secondly,
$Sp(2,4) \supset Sp(2)_1 \times Sp(2)_2 \times Sp(2)_3 \sim SU(2)_1 \times SU(2)_2 \times SU(2)_3$
and $(A_{+a}, -A_{-a}^*)$ is a doublet of $SU(2)_a$. One could define three quaternions
\beq
\label{conf17}
Q_a = \left( \begin{array}{cc}  A_{+a} & A_{-a} \\ -A_{-a}^* & A_{+a}^*  \end{array} \right)
\qquad\qquad a=1,2,3
\eeq
with a left action of $SU(2)_a$ and a right action of the superconformal $SU(2)_R$. 
They verify (for each $a$)
\beq
\label{conf17a}
Q_a \, Q_a^\dagger  = Q^\dagger_a \, Q_a = \det Q_a \, I_2,
\qquad\qquad \det Q_a = |A_{+a}|^2 + |A_{-a}|^2.
\eeq

The second condition (\ref{conf11}) from $N=2$ supergravity becomes:
\beq
\label{conf19}
\vec A_+^* \cdot \vec A_+ + \vec A_-^* \cdot \vec A_- = -{1\over\kappa^2}, \qquad\qquad
\vec A^*\cdot\vec A = \vec A^\dagger \eta \vec A = -|A_1|^2 + |A_2|^2 - |A_3|^2 .
\eeq
With Eq.~(\ref{Tis2}), condition (\ref{conf12}) leads to three (real) equations:
\beq
\label{conf22}
\begin{array}{rcl}
\vec A_+^\dagger \,i\eta \hat T\, \vec A_+ &=&  \vec A_-^\dagger \,i\eta \hat T\, \vec A_-   ,
\crbig
\vec A_-^\dagger \,i\eta \hat T \, \vec A_+  &=& 0  
\end{array}
\eeq
($[i\eta \hat T]^\dagger = i\eta\hat T$). 
With the explicit form of $\hat T$, Eq.~(\ref{Tis1}), and defining dimensionless fields 
$a_{\pm i} = \sqrt2\kappa A_{\pm i}$,
the four constraints (\ref{conf19}) and (\ref{conf22}) read finally
\beq
\label{sol10}
\begin{array}{ll}
I:\qquad&
|a_{+1}|^2 + |a_{-1}|^2 - |a_{+2}|^2 - |a_{-2}|^2 + |a_{+3}|^2 + |a_{-3}|^2 = 2,
\crbig
II:\qquad&
-|a_{+1}|^2 + |a_{+2}|^2 - |a_{+3}|^2 - \chi |a_{+1}-a_{+2}|^2
\crbig &\hspace{2.1cm}
= -|a_{-1}|^2 + |a_{-2}|^2 - |a_{-3}|^2 - \chi |a_{-1}-a_{-2}|^2,
\crbig
III:\qquad&
0 = -a_{+1}\ov a_{-1} + a_{+2}\ov a_{-2} - a_{+3}\ov a_{-3}
- \chi (a_{+1}-a_{+2})(\ov a_{-1} - \ov a_{-2}) .
\end{array}
\eeq
They are invariant under Heisenberg variations (\ref{Heisvar}) of $\vec a_+$ and $\vec a_-$.
The case $\chi=0$ has been considered by Galicki \cite{G1}. Since it leads to 
$SU(1,2)/SU(2)\times U(1)$, coordinates more appropriate for this larger isometry have been used.

\subsection{Solving the constraints}\label{secconst}

To solve the constraints (\ref{sol10}), we insist on keeping in $\vec a_-$ a field $\Phi$ which 
transforms under the Heisenberg variations\footnote{See Eq.~(\ref{Heisvar}).}
$\delta_H\, \vec a_- = (\alpha X+\beta Y+ \gamma Z)\,\vec a_-$ with a complex shift:
\beq
\label{sol4}
\delta_H \, \Phi = \alpha - i \beta.
\eeq
This is the case if $a_{-1}=a_{-2}$, and $a_{-3}$ is then invariant. We may define 
$\ov\Phi = a_{-1}/a_{-3}$ and constraint $III$ reduces to $a_{+3}= (a_{+2}-a_{+1})\Phi$.
Since 
$$
\delta_H \left({a_{+2} + a_{+1} \over a_{+2} - a_{+1} }\right) = - 2i\gamma 
+ 2(\alpha + i \beta){ a_{+3} \over a_{+2} - a_{+1}}
= -2i\gamma + 2 \Phi\, \delta_H\ov\Phi,
$$
we finally define 
\beq
\label{sol5}
S =  {a_{+2} + a_{+1} \over a_{+2} - a_{+1}} + Y , \qquad\qquad
\delta_HS = -2i\gamma + 2(\alpha+i\beta)\Phi
\eeq 
and the quantity
\beq
\label{sol5b}
Y = S+\ov S - 2 \Phi\ov\Phi
\eeq 
is invariant under Heisenberg variations. The algebra follows from
$[\delta_H^\prime,\delta_H] = (\alpha^\prime\beta-\alpha\beta^\prime)[X,Y]
= 2(\alpha^\prime\beta-\alpha\beta^\prime)Z$:
$$
[\delta_H^\prime, \delta_H]S = 2(\alpha^\prime+i\beta^\prime) \delta_H\Phi 
- 2(\alpha+i\beta) \delta_H^\prime\Phi
= -4i(\alpha^\prime\beta-\alpha\beta^\prime)
= 2(\alpha^\prime\beta-\alpha\beta^\prime)Z.
$$
These definitions are summarized in the choice
\beq
\label{sol11b}
\vec a_- = {K\over\Delta} \left( \begin{array}{c} \ov\Phi \\ \ov\Phi \\ 1 \end{array}\right),
\qquad\qquad
\vec a_+ = {1\over\Delta}\left( \begin{array}{c} S-Y-1 \\ S-Y+1 \\  a \end{array} \right),
\eeq
with complex fields $S$, $\Phi$ and $a$.
The four available gauge choices have been used to take $\Delta=|\Delta|$, $K=|K|$ 
and $a_{-1} = a_{-2}$.
Under Heisenberg variations, $\Delta$ and $K$ are invariant.
Hence, we are left with eight real scalar fields submitted to the four constraints (\ref{sol10}) 
which drastically simplify:
\beq
\label{sol12}
\begin{array}{ll}
I:\qquad&
\Delta^2\Bigl(2 - |a_{+1}|^2 + |a_{+2}|^2 - |a_{+3}|^2 \Bigr)  = K^2,
\crbig
II:\qquad&
2(S+\ov S) - |a|^2 - 4 Y= 4\chi - K^2 ,
\crbig
III:\qquad&
a  = 2\Phi.
\end{array}
\eeq
Hence, the solution is
\beq
\label{sol16}
\vec a_- = \sqrt{Y + 2 \chi \over Y + \chi}
 \left( \begin{array}{c} \ov\Phi \\ \ov\Phi \\ 1 \end{array}\right),
\quad\qquad
\vec a_+ ={1\over\sqrt{2(Y + \chi)}}
\left( \begin{array}{c} S - Y - 1 \\ S- Y + 1 \\ 2\Phi \end{array} \right).
\eeq
The solution implies $Y+\chi>0$ if $\chi>0$ or $Y+2\chi>0$ if $\chi<0$.
The scalar kinetic Lagrangian (\ref{conf15}) obtained from this solution is\footnote{
All fields and parameter $\chi$ are dimensionless.}
\beq
\label{sol18a}
\begin{array}{rcl} 
\kappa^2 {\cal L} &=& \displaystyle
{ (Y+3\chi) \over 4(Y+2\chi)(Y+\chi)^2}(\partial_\mu Y)^2 - {2\over Y+\chi} \, \partial_\mu\Phi\,
\partial^\mu\ov\Phi
\crbig
&& \displaystyle
+ {1\over2(Y+\chi)(Y+3\chi)} \, \left[ \Im(\partial_\mu S - 2\ov\Phi\, \partial_\mu\Phi)\right]^2 
\crbig
&& \displaystyle
+ {1 \over 2(Y+\chi)^2} \, \left[ \Im(\partial_\mu S - 2\ov\Phi\, \partial_\mu\Phi)\right]^2
+ {4(Y+2\chi)\over (Y+\chi)^2} \, \partial_\mu\Phi\,\partial^\mu\ov\Phi. 
\end{array}
\eeq
The first line comes from the basic scalar kinetic terms ${\cal L}_{kin.}$ in Lagrangian
(\ref{conf15}). The second line is the contribution ${\cal L}_T$ of the 
gauge field of $T$, the third line arises from the supergravity $SU(2)_R$ gauge fields. Each term is 
separately invariant under Heisenberg variations.
Collecting terms, the final form of the theory is
\beq
\label{sol18b}
\begin{array}{rcl}
\kappa^2 {\cal L} &=& \displaystyle
{Y+3\chi \over (Y+\chi)^2 } \left[ {1\over4} {(\partial_\mu Y)^2 \over Y+2\chi}
+ 2 \partial_\mu\Phi \, \partial^\mu\ov\Phi\right]
\crbig
&& \displaystyle
+ {Y+2\chi \over (Y+3\chi)(Y+\chi)^2 } \Bigl( \partial_\mu \Im\hat S 
- 4 \Re\Phi\, \partial_\mu\Im\Phi \Bigr)^2,
\end{array}
\eeq
where 
\beq
\label{newS}
\hat S = S+\Phi^2,
\eeq
for which $Y=\hat S + \ov{\hat S} - (\Phi+\ov\Phi)^2$ and 
$\Im(dS-2\ov\Phi \,d\Phi) = d\Im\hat S - 4\Re\Phi\, d\Im\Phi$. 
From the existence of solutions (\ref{sol16}) and positivity of the
Lagrangian, the range of $Y$ is 
$Y+\chi>0$ if $\chi>0$ and $Y+3\chi>0$ if $\chi<0$
Writing as usual
\beq
\label{Lmetric}
{\cal L} = {1\over\kappa^2} \, g_{ab} (\partial_\mu q^a)(\partial^\mu q^b)
= G_{ab}(\partial_\mu q^a)(\partial^\mu q^b),
\eeq
$q^a=(Y, \Re\Phi,\Im \Phi, \Im\hat S)$, and comparing $ds^2= g_{ab} \,dq^a dq^b$ with 
expression (\ref{CP2}), we see that the hypermultiplet kinetic metric $g_{ab}$ is the CPH metric with
\beq
\label{sol19}
Y = V-2\chi = \rho^2-2\chi,
\eeq
and with\footnote{This choice is not unique. We may for instance rotate $\Phi$ using isometry $M$.}
\beq
\label{sol20}
\Phi = {1\over\sqrt2} (\eta+i\varphi), \qquad\qquad
\Im\hat S = - 2 \tau.
\eeq
Positivity of kinetic terms is obtained if $V= \rho^2 >|\chi|$ which is, as explained at the end
of Subsection \ref{secCPH}, the natural domain of $V$. 

As already observed, the case $\chi=0$ corresponds to 
the $SU(2,1)/SU(2)\times U(1)$ metric
\beq
\label{sol8c}
ds^2 = {1\over Y^2}\left[ {1\over4}dY^2 + \Bigl(d\Im\hat S-4\Re\Phi \,d\Im\Phi\Bigr)^2 \right] 
+ {2\over Y}\, d\Phi d\ov\Phi .
\eeq
With K\"ahler coordinates $\hat S$ and $\Phi$, the K\"ahler potential is $K=-\ln Y$, with 
$Y=V=\hat S+\ov{\hat S} - (\Phi+\ov\Phi)^2$.

This relatively simple construction of the one-loop-corrected dilaton hypermultiplet metric allows easily
to derive the full $N=2$ supergravity Lagrangian, using $N=2$ superconformal tensor calculus
\cite{N=2conf, G1, G2}.

\section{Zero-curvature hyper-K\"ahler limit}\label{seclimit}
\setcounter{equation}{0}

All quaternion-K\"ahler metrics are Einstein spaces with nonzero curvature. With one hypermultiplet,
the scalar kinetic Lagrangian (\ref{Lmetric}) verifies \cite{BW}
\beq
\label{Ricci1}
R_{ab} = -6\,g_{ab} = -6 \kappa^2\, G_{ab}.
\eeq
The link with global $N=2$ supersymmetry is realized by defining a $\kappa\rightarrow0$ 
hyper-K\"ahler limit of the CPH metric (\ref{CP2}) or (\ref{sol18b}) in which, if feasible, the Heisenberg 
algebra does not contract to an abelian symmetry. 
As observed in Subsection \ref{secCPH}, the magnitude of $\chi$ can be eliminated by rescaling 
of the coordinates (in the absence of D-branes). We then have three $|\chi|$-independent cases to examine: firstly, positive 
$\chi$, with $V>0$; secondly, $\chi=0$ ($V>0$) which is $SU(1,2)/SU(2)\times U(1)$; thirdly, a 
negative $\chi$, with $V > |\chi|$. In each case, we should seek to find a parameter-free 
zero-curvature limit. The most interesting case turns out to be $\chi$ negative, which we first study. 

With $\chi$ negative, we are interested in the CPH metric in the region $V+\chi\sim0$.
We then apply to metric (\ref{CP2}) the following change of variables:
\beq
\label{change}
\begin{array}{rclrcl}
V &=& 2|\chi| \, \kappa^{2/3} \mu^{-1/3}\,C- \chi \,, \qquad&\qquad
\varphi&=& \sqrt{|\chi|} \,\kappa^{2/3} \mu^{-1/3}\,\hat{\varphi}\,,
\crbig
\eta&=&\sqrt{|\chi|} \,\kappa^{2/3} \mu^{-1/3}\,\hat{\eta}\,,\qquad&\qquad
\tau&=&|\chi|\,\kappa^{4/3} \mu^{1/3}\,\hat{\tau}\,,
\end{array}
\eeq
where $\mu$ is an arbitrary mass scale. Positivity of the metric, $V+\chi>0$ implies $C>0$. 
While the original fields 
are dimensionless, the new, hatted, fields $(C, \hat \phi,\hat\eta,\hat\tau)$
have canonical dimension. With this choice of dependence in $\kappa$, the resulting metric is
\beq
\label{flat4}
\begin{array}{rcl} 
ds^2 \,\,=\,\, g_{ab} \, dq^adq^b&=& \displaystyle
{\kappa^2\over2} {\mu C \over \bigl[(\kappa\mu)^{2/3}C 
+ \mu \bigr]^2 } \left[  
{dC^2 \over 2\kappa^{2/3}\mu^{-1/3}C + 1} + d\hat\eta^2 + d\hat\varphi^2 \right]
\crbig
&& \displaystyle
+ {\kappa^2\mu^2 \over 2C} {2(\kappa\mu)^{2/3}C + \mu \over
[(\kappa\mu)^{2/3}C + \mu ]^2 }  
\left[ d\hat\tau + \frac{1}{\mu}\hat\eta d\hat\varphi \right]^2,
\end{array}
\eeq
since $\chi = -|\chi|$.
Using this metric in Lagrangian (\ref{Lmetric}), the overall factor $\kappa^2$ cancels and we can 
take the limit $\kappa\rightarrow0$, with result
\beq
\label{flat5}
{\cal L}_{\kappa\rightarrow0} = {C \over 2\mu}
\left[ (\partial_\mu C)^2  + (\partial_\mu \hat{\eta})^2 + (\partial_\mu \hat{\varphi})^2\right]
+ {\mu\over 2C}  \left[\partial_\mu\hat{\tau} + {1\over\mu}\hat{\eta}\,\partial_\mu\hat{\varphi}\right]^2\,.
\eeq
This scalar Lagrangian has the hyper-K\"ahler metric with Heisenberg symmetry (\ref{HK5}) with $A=1/\mu$ and $B=0$ and with relations $\Phi = {1\over\sqrt2}(\hat\eta + i \hat \varphi)$, $\hat\tau=2\tau$.
As noticed earlier, parameter $B$ can always be absorbed in a shift of $C$, as long as 
$A\ne0$.

Notice that to obtain limit (\ref{flat5}), we only need the change of variables (\ref{change}) 
up to higher orders in $\kappa$. In particular, according to Eq.~(\ref{dilaton}), we
may write the four-dimensional string dilaton as
\beq
\label{gslimit}
\begin{array}{rcl}
e^{-2\phi_4} &=& 2|\chi|\kappa^{2/3} \mu^{-1/3}\,C - 2\chi ,
\crbig
\phi_4 &=& \langle\phi_4\rangle - \kappa^{2/3}\mu^{-1/3} \hat\phi_4 ,
\crbig
e^{-2\langle\phi_4\rangle} &=& -2\chi \,\,=\,\, 2|\chi|,
\qquad\qquad\qquad   C \,\, = \,\, 2\hat\phi_4 ,
\end{array}
\eeq
in terms of the fluctuation $\hat\phi_4$ and of the background value $\langle\phi_4\rangle$.
Since $|\chi| = \chi_1 = \chi_E/(12\pi)$, we are considering the case of a positive Euler number
$\chi_E=2(h_{11}-h_{21})$, with $h_{11},h_{12}$ the corresponding Betti numbers of the CY$_3$ manifold. A typical example with a single hypermultiplet would be IIA strings on a CY$_3$ manifold with $h_{21}=0$.
Positivity-related questions with several hypermultiplets, as is in particular 
the case with a negative Euler number, should be reanalyzed. 

Comparing the scalings (\ref{change}) and the identification of the string
coupling in the last Eq. (\ref{gslimit}), we see that the R-R fields
$\eta$ and $\varphi$ carry as expected a supplementrary factor $g_{string}$.

We could also consider the single-tensor version of the theory. 
Dualizing $\hat\tau$ into $H_{\mu\nu\rho}$, we find
\beq
\label{flat6}
\begin{array}{rcl}
{\cal L}_{\kappa\rightarrow0,ST} &=& \displaystyle
{C \over \mu }\left[ {1\over2}(\partial_\mu C)^2
+ {1\over12} H^{\mu\nu\rho} H_{\mu\nu\rho} 
+ (\partial_\mu\ov\Phi)(\partial^\mu\Phi) \right]
\crbig
&& \displaystyle
- {i\over 12\mu} \epsilon^{\mu\nu\rho\sigma} ( \ov\Phi\partial_\mu\Phi - \Phi\partial_\mu\ov\Phi )
 H_{\nu\rho\sigma} .
\end{array}
\eeq
This is the bosonic sector (\ref{Hsymbos}) of the single-tensor theory 
(\ref{Hsym5}) with again $A=1/\mu$ and $B=0$. Then, for negative $\chi$,
the $N=2$ supergravity hypermultiplet with Heisenberg symmetry is 
described in the global supersymmetry limit by the unique nontrivial
theory with the same symmetry. 

For completeness, we may also consider the case of the CPH metric with 
positive $\chi$. The interesting limiting regions are $V\sim0$ and 
$V-\chi\sim 0$. If $V = \rho^2 \ll \chi$, 
\beq
ds^2_{CPH} = {1\over\chi}( d\rho^2 + d\eta^2 + d\varphi^2) + {4\rho^2\over\chi^3}
(d\tau + \eta \,d\varphi)^2.
\eeq
The appropriate rescalings are $(\rho,\eta,\varphi,\tau) =
(\sqrt\chi\kappa\hat\rho,\sqrt\chi\kappa\hat\eta,\sqrt\chi\kappa\hat\varphi,\chi\hat\tau)$ to obtain
\beq
ds^2_{CPH} = \kappa^2 \left[ d\hat\rho^2 + d\hat\eta^2 + d\hat\varphi^2 
+ 4\hat\rho^2 (d\hat\tau + \kappa^2\,\hat\eta d\hat\varphi)^2 \right].
\eeq
The Heisenberg symmetry acting on the rescaled fields has algebra $[X,Y]=2 \kappa^2 Z$.
In the limit $\kappa\rightarrow0$, it contracts to $[X,Y]=0$ and we find
\beq
\lim_{\kappa\rightarrow0} \, {1\over\kappa^2} ds^2_{CPH} = d\hat\rho^2 + 4\hat\rho^2d\hat\tau^2 + d\hat\eta^2 + d\hat\varphi^2,
\eeq
which is the trivial four-dimensional euclidean space.  
The second region of interest if $\chi>0$ is $V-\chi\sim0$. First, we change coordinates to
\beq
V= 2\lambda C + \chi, \qquad \eta= \lambda \hat\eta/\sqrt\chi,
\qquad \varphi = \lambda\hat\varphi / \sqrt\chi, 
\qquad \tau =\lambda \hat\tau
\eeq
and the metric for $\lambda\rightarrow0$ and $\chi$ finite reads 
\beq
ds^2_{CPH}  = {1\over 2C^2} \left[ dC^2
+ d\hat\eta^2 + d\hat\varphi^2 + d\hat\tau^2 \right].
\eeq
This limiting metric is $SO(1,4)/SO(4)$, again with $R_{ij}=-6g_{ij}$ and with radius 
$\sim\langle C \rangle$. In the large radius, zero-curvature limit, the metric is trivial.
Finally, in the $SU(1,2)/SU(2)\times U(1)$ case $\chi=0$, the zero-curvature limit is again trivial.

The conclusion is that in the zero-curvature limit, the CPH one-loop Lagrangian for the 
dilaton hypermultiplet is the hyper-K\"ahler $N=2$ sigma-model with Heisenberg symmetry
(\ref{Hsym5}). If the one-loop parameter $\chi$ is negative, then $A\ne0$ and the Heisenberg 
algebra has a non-trivial realization in this limit. If $\chi\ge0$ however, $A=0$ and the
limit of $N=2$ global supersymmetry is the free hypermultiplet. In the string context, the above non-trivial limit can be taken if the string coupling is tuned at a fixed value, according to the third line of Eq.~(\ref{gslimit}), which applies with positive Euler number.

In a recent paper \cite{AADT}, we have constructed the interaction of a 
hypermultiplet with the Dirac-Born-Infeld Maxwell Lagrangian. The hypermultiplet sector has a full linear $N=2$ supersymmetry while the second supersymmetry is nonlinearly realized on the Maxwell superfield $W_\alpha$.
As an application of our results, we can easily use our identification
of the string universal hypermultiplet. The bosonic DBI action, after
elimination of the Maxwell auxiliary field and using the single-tensor formulation, is\footnote{In Ref. \cite{AADT}, this is the electric version of the theory, induced by a $N=2$ Chern-Simons coupling $g B\wedge F$.}
\beq
\label{DBI}
\begin{array}{rcl}
{\cal L}_{DBI} &=& \displaystyle
{1\over 8 {\mathf f}}(2g \textrm{Re} \Phi- {1\over {\mathf f}})\left[ 1-\sqrt{1+{2g^2C^2\over(2g\Re\Phi- {1\over{\mathf f}} )^2}}\sqrt{-\det(\eta_{\mu\nu}+2\sqrt{2}{\mathf f}F_{\mu\nu})} \right]
\crbig
&+&\displaystyle g \epsilon^{\mu\nu\rho\sigma}\left({{\mathf f}\over 4} \textrm{Im} \Phi F_{\mu\nu}F_{\rho\sigma}-{1\over 4} B_{\mu\nu}F_{\rho\sigma}+{1\over 24{\mathf f}}C_{\mu\nu\rho\sigma}\right).
\end{array}
\eeq
In this expression, ${\mathf f}$ is the breaking scale of the second,
nonlinearly realized supersymetry (with dimension (energy)$^{-2}$) and
$g$ is the Chern-Simons coupling\footnote{In contrast to Ref. \cite{AADT}, we have defined single-tensor fields with canonical dimension
so that $g$ has dimension (energy). We also chose the Fayet-Iliopoulos term to be $1/{\cal F}$ so that gauge kinetic terms are canonically normalized at
$\Re\Phi=0$.} (equal to the string coupling for a D3-brane).
The four-form field 
$C_{\mu\nu\rho\sigma}$ is a component of the single-tensor multiplet required by supersymmetry of the nonlinear theory \cite{AADT}.

Since we have control of the kinetic Lagrangian of the universal string hypermultiplet in the global supersymmetry limit, we can then identify 
the single-tensor fields in terms of string fields. First, $C$ is the global
dilaton and  $B_{\mu\nu}$ is the NS-NS tensor. 
Then, the complex scalar $\Phi$ includes the R-R fields. The
supersymmetric minimum of the scalar potential included in theory (\ref{DBI})
implies $\langle C\rangle =0$ and $\Phi$ corresponds to 
flat directions of this vacuum. 

\section{Conclusions}\label{secfin}
\setcounter{equation}{0}

In this work, we analyzed the effective field theory of the universal dilaton hypermultiplet of type II string compactifications on a CY$_3$ manifold with 
a special emphasis on the global supersymmetry limit. The perturbative isometries form the two-dimensional Euclidean algebra $E_2$ with a central extension, which contains a Heisenberg subalgebra. Using this isometry as a guiding principle and the method of quaternionic quotient in conformal supergravity, we rederived the two-derivative $N=2$ supergravity action, depending on a deformation parameter that corresponds to the one-loop correction proportional to the Euler number of the CY$_3$ manifold. We then established the precise connection with a one-parameter family of 
hyper-K\"ahler spaces in $N=2$ global supersymmetry, possessing the same isometry, by defining a non-trivial gravity decoupled limit characterized by a new mass scale. This requires the string coupling to be tuned at a fixed value which only occurs for a positive Euler number. As the latter becomes large, the theory becomes weakly coupled, justifying the perturbative approximation in taking the global limit. Notice that in the absence of moduli stabilization effects, this positivity requirement is compatible with the possibility of choosing 
$h_{21}=0$ in type IIA, that guarantees the absence of other hypermultiplets in the spectrum, which could modify the constraints we derived from the positivity of the metric. It would be interesting to understand this requirement in the general case.

\section*{Acknowledgements}

This work has been supported by the Swiss National Science Foundation. 
The work of I.A. and P.T. was supported in part by the European Commission under the ERC 
Advanced Grant 226371 and the contract PITN-GA-2009-237920. I.A. was also supported by 
the CNRS grant GRC APIC PICS 3747. P.T. would like to thank the `Propondis' Foundation for its support as well as the Galileo Galilei Institute for Theoretical Physics for
the hospitality and the INFN for partial support during the early stage of this work.


\end{document}